\documentclass[aps,prb,twocolumn,superscriptaddress, show pacs]{revtex4-1}  
\usepackage{bm, amsmath}
\usepackage{graphicx}
\usepackage{color}
\usepackage{epstopdf}
\usepackage{graphicx}
\usepackage{dcolumn}
\usepackage{bm}
\usepackage{subfigure}

\begin{document}

\title{Field and anisotropy driven transformations of spin spirals in cubic skyrmion hosts}

\author{A. O. Leonov}
\thanks{leonov@hiroshima-u.ac.jp}
\affiliation{Department of Chemistry, Faculty of Science, Hiroshima University Kagamiyama, Higashi Hiroshima, Hiroshima 739-8526, Japan}
\affiliation{IFW Dresden, Postfach 270016, D-01171 Dresden, Germany} 

\author{C. Pappas}
\thanks{c.pappas@tudelft.nl}
\affiliation{Faculty of Applied Sciences, Delft University of Technology, Mekelweg 15, 2629JB Delft, The Netherlands}

\author{I. K\'ezsm\'arki}
\thanks{istvan.kezsmarki@physik.uni-augsburg.de}
\affiliation{Experimental Physics V, Center for Electronic Correlations and Magnetism,
University of Augsburg, 86135 Augsburg, Germany}

\date{\today}

\begin{abstract}
{We discuss distinctive features of spiral states in bulk chiral magnets such as MnSi and Cu$_2$OSeO$_3$ that stem from the effect of the cubic magneto-crystalline anisotropy.  
First of all, at both the helical-to-conical and the conical-to-ferromagnetic transitions, taking place at H$_{c1}$ and H$_{c2}$, respectively,
the cubic anisotropy leads to reversible or irreversible jump-like reorientations of the spiral wavevectors.
%
The subtle interplay between the easy and hard anisotropy axes  gives rise to a phase transition between elliptically distorted conical states almost without any detectable change in the period. 
We show that the competition between on-site cubic and exchange anisotropy terms can also lead to oblique spiral states. 
Our work gives clear directions for further experimental studies to reveal
 theoretically predicted spiral states in cubic helimagnets beyond the aforementioned well-established states thus, can help to understand the magnetic phase diagram of these archetypal skyrmion hosts.  
In addition, we show that properties of isolated skyrmions such as inter-skyrmion attraction, orientation and/or nucleation are also rooted in the properties of host spirals states, in which skyrmions are stabilized.
}
\end{abstract}

         
\maketitle

\section{Introduction}

Following the pioneering theoretical works of Dzyaloshinskii  \cite{Dz64,ishikawa} and the experimental findings in the 1980's, where the magnetic phase diagram of MnSi  has been studied by different techniques and the results were summarized by Kadowaki \cite{Kadowaki82},  chiral helimagnetism \cite{Izyumov} experiences an increasing interest over the last years. This is mainly due to the 
the discovery of skyrmions \cite{Muehlbauer09,yuFeCoSi,YuFeGe} -- two-dimensional localized whirls of the magnetization, which may  be used in future information storage and data processing devices.  
%
%
Presently the field of skyrmionics encompasses activities  to create real prototypes of skyrmion-based devices in nanosystems such as nanolayers \cite{yuFeCoSi,YuFeGe}, nanowires \cite{Liang} or nanodots \cite{nanodiscs,Du2015}, exploiting the  small size, topological protection and the ease with which skyrmions can be manipulated by electric currents.
This swerve from bulk  cubic helimagnets  to nanosystems, witnessed over the last years, was stipulated by the reduced skyrmion stability in the former case, where skyrmions appear forming hexagonal lattices in a small pocket of the temperature-magnetic field phase diagram just below the transition temperature $T_C$, the so-called A-phase.

Recent experimental findings, however,  revealed a  vast area of low-temperature skyrmion stability  in the bulk insulating cubic helimagnet Cu${_2}$OSeO${_3}$, when the magnetic field is applied along the easy $\langle$001$\rangle$ crystallographic axis  \cite{qian2018,chacon2018}. 
Remarkably, the theoretical explanation of this phenomenon 
is mainly based on the effect of single-ion cubic anisotropy on one-dimensional spiral states \cite{Bannenberg2019,droplets}. 
In fact, the ideal magnetization rotation in the conical phase, the main competitor of skyrmions, can be impaired by the easy and hard anisotropy axes for specific directions of the magnetic field. 
Through this mechanism, skyrmions, which are more resilient to  anisotropy-induced deformations, due to their two-dimensional nature, gain stability over extended regions of the phase diagram.
In general, it was shown that the cubic anisotropy has a direct impact on the spiral states and thus indirectly is at the origin of many remarkable phenomena for skyrmions. 

The experimental findings provide hints that the cubic anisotropy in Cu${_2}$OSeO${_3}$ initiates the nucleation of thermodynamically stable skyrmions via two different mechanisms\cite{Bannenberg2019}:  either via rupture formation of \textit{metastable} spiral states with wave vectors perpendicular to the field, or 
via torons \cite{leonov2018x}. The latter are localized particles consisting of two Bloch points at finite distance and a convex-shaped skyrmion stretching between them. In   Cu${_2}$OSeO${_3}$ they can be generated by the inhomogeneous magnetic environment provided by the smooth transition between tilted spiral domains, which appear when the magnetic field is applied along the easy $\langle$001$\rangle$ crystallographic axis \cite{qian2018}.  
This tilted spiral phase constitutes a major deviation from the generic phase diagram of cubic chiral magnets \cite{Bauer2016, Bauer2017} and is stabilized by a competition between on-site cubic and exchange anisotropies. This new phase may also give rise to new topological magnetic defects, such as isolated skyrmions, with interesting static and dynamic properties. 
Moreover, versatile skyrmion phases have been recently identified both at and above room temperature in a family of cubic chiral magnets: $\beta$-Mn-type Co-Zn-Mn alloys with a different chiral space group from that of B20 compounds \cite{Tokunaga2015}. 

Here we undertake a systematic study of the effects imposed by  cubic anisotropy on one-dimensional spiral states.
First, we show that cubic anisotropy  accounts for the reorientation processes of spiral states that occur not only at the low critical field $H_{c1}$,  but also near the transition into the saturated state $H_{c2}$, an effect that  might at a first sight seem counter-intuitive. 
Furthermore, we show that  the existence of easy and hard anisotropy axes  distorts elliptically the magnetization    rotation and this effect may stabilise   new spiral phases with no changes of the spiral orientation or  period, but with  detectable transitions between them.
Our theoretical approach explains the behavior  of the bulk helimagnets Cu${_2}$OSeO${_3}$, or MnSi, and provides a mechanism through which it is  possible to generate  thermodynamically stable skyrmionic spin textures over large fractions of their phase diagrams. 
%

\section{Phenomenological theory of spiral states in cubic helimagnets \label{PTBH}}

\begin{figure}
\includegraphics[width=0.99\columnwidth]{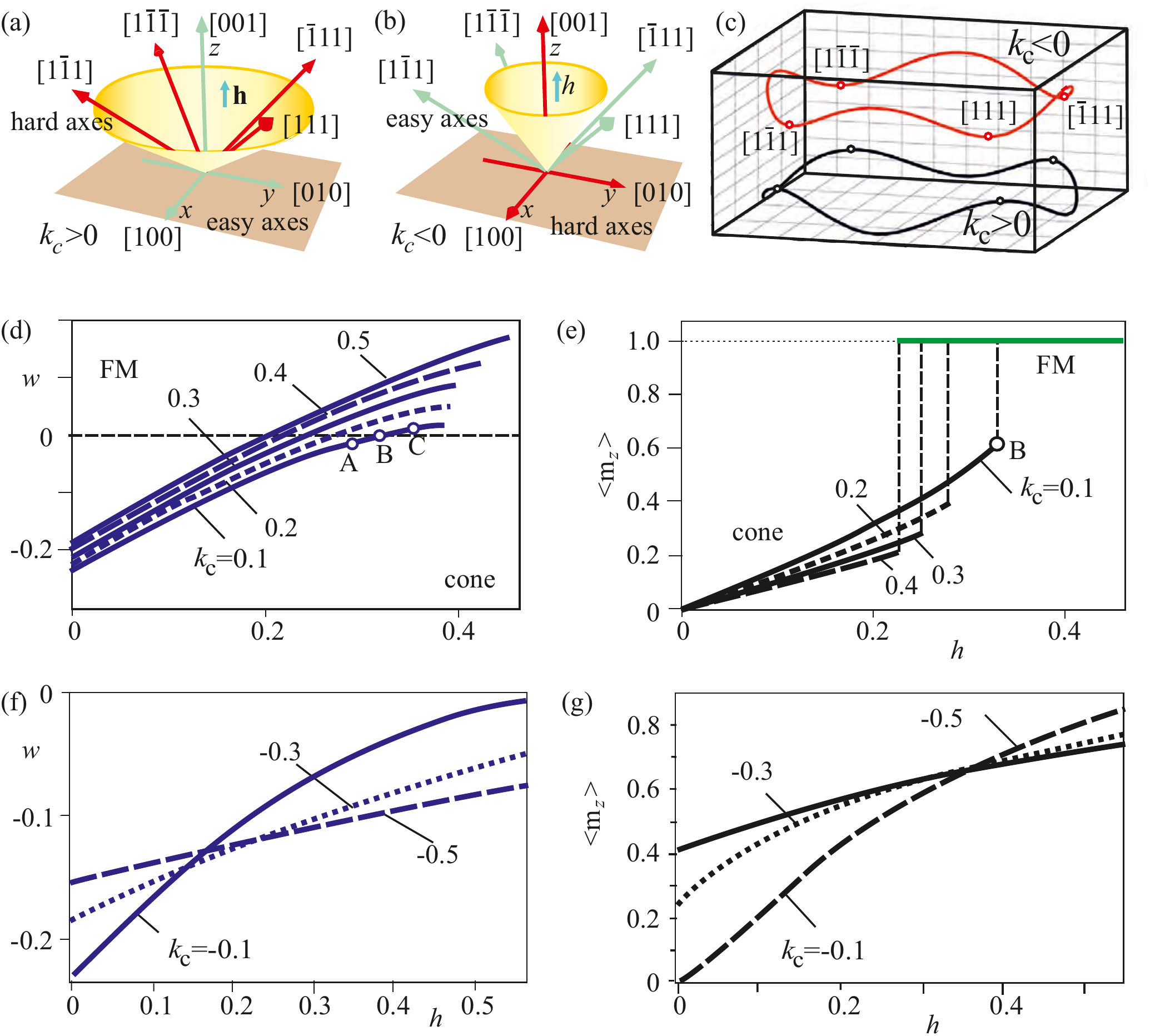}
\caption{
(color online)  Schematic representations of the magnetization rotation in the conical phase in the presence of cubic anisotropy with $k_c>0$ (a) and $k_c<0$ (b)  shown together with the magnetization traces in a space (c). Depending on the orientation of the cone propagation direction ($z$) with respect to the easy (green arrows) and hard (red arrows) anisotropy axes and the applied magnetic field ($h||[001]$) the energy density of the conical pahse can be increased or  reduced. 
This phase undergoes a first-order phase transition into the ferromagnetic state at the critical field $h_{c2}$ for $k_c>0$ and $\mathbf{h}||[001]$. This is seen at the energy density difference between the conical and ferromagnetic phases, depicted in (d), and the  magnetization curves shown in (e). The jumps in (e) occur at the field value where the energy density difference in (d) becomes zero, e.g., in the point B for $k_c=0.1$. For $k_c<0$ and $\mathbf{h}||[001]$ (f),(g)  the mutual arrangement of easy and hard anisotropy axes,  shown in (b)  underlies a second-order phase transition.
\label{Fig1}
}
\end{figure}

\subsection{The general micromagnetic energy functional}

Within the phenomenological theory introduced by Dzyaloshinskii \cite{Dz64} the magnetic energy density of a  non-centrosymmetric ferromagnet with spatially dependent magnetization $\mathbf{M}$ can be written as 
\begin{equation}
W(\mathbf{M})=\underbrace{A \sum_{i,j}\left(\frac{\partial m_j}{\partial x_i}\right)^2
+D\,w_D(\mathbf{M}) -\mathbf{M}\cdot\mathbf{H}}_{W_0(\mathbf{M})} + W_a(\mathbf{m})
\label{DMdens1}
\end{equation}
where $A$ and $D$ are coefficients  of exchange and Dzyaloshinskii-Moriya interactions (DMI); $\mathbf{H}$ is an  applied magnetic field; $x_i$ are the Cartesian components of the spatial variable. 
$w_D$ is composed of Lifshitz invariants $\mathcal{L}^{(k)}_{i,j} = M_i \partial M_j/\partial x_k - M_j  \partial M_i/\partial x_k$ that are energy terms involving first derivatives of the magnetization  with respect to the spatial coordinates. In the following, all calculations  will be done for cubic helimagnets with $w_D=\mathbf{m}\cdot\nabla\times\mathbf{m}$ although the results may be applied for magnets with other symmetry classes \cite{JETP89} including different combinations of Lifshitz invariants. 
%
%
$W_0(\mathbf{M})$ includes only basic interactions essential to stabilize skyrmion and helical states and specifies their most general features attributed to all chiral ferromagnets. 
Next we discuss the physics governed by these dominant interactions. The effects of anisotropy $W_a(\mathbf{m})$, usually representing a smaller energy scale, will be discussed later.

\begin{figure*}
\includegraphics[width=2.0\columnwidth]{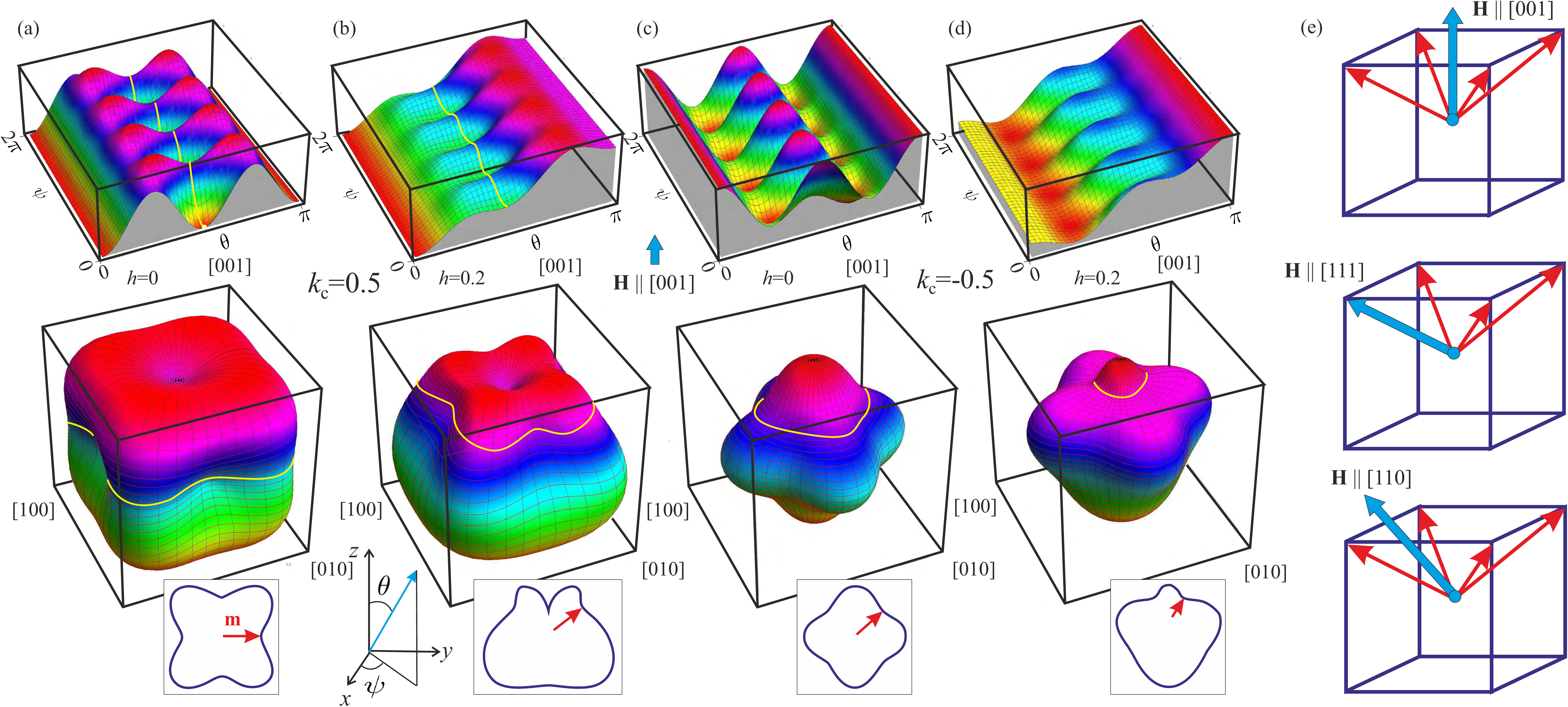}
\caption{
(color online)  Cubic anisotropy energy density (\ref{additional})  as a function of the polar and azimuthal angles $\theta$ and $\psi$ of the ferromagnetic state shown as two-dimensional surfaces  (first row) and  three-dimensional polar plots  (second row).    The graphs in the third row depict cuts of the corresponding polar plots with $\psi=0$. The iath of the rotating magnetization in the conical phase is shown by the yellow lines.  The field is applied along the [001] direction. (a) and (b) correspond to $k_c=0.5$ and in this case, the states of the magnetization in the cone correspond only to the local minima. (c) and (d) correspond to $k_c=-0.5$ and the magnetization rotates to sweep the global minima of the cubic anisotropy. (e) Re-orientation of the spirals into the conical state  for $k_c < 0$ and different directions of the field. $\mathbf{q}$-vectors of spirals are shown by red arrows and point along $\langle111\rangle$ crystallographic axes (see text for details).
\label{Fig2}
}
\end{figure*}

\subsection{One-dimensional spiral modulations within the isotropic theory}

In chiral magnets the Dzyaloshinskii-Moriya interactions $w_{D}$  play a crucial role in destabilizing the homogeneous ferromagnetic arrangement and twisting it into a helix. At zero magnetic field the helices are single-harmonic modes forming the global minimum of the functional $W_0(\mathbf{M})$ \cite{Dz64}, $\mathbf{M}= M_s \left[ \mathbf{n}_1 \cos
\left(\mathbf{q} \cdot \mathbf{r} \right)+ \mathbf{n}_2 \sin  \left(\mathbf{q}\cdot \mathbf{r} \right) \right]$. 
$\mathbf{n}_1$, $\mathbf{n}_2$  are the unit vectors in the plane of the magnetization rotation orthogonal to the wave vector $\mathbf{q} = 1/(2L_D)$ ($\mathbf{n}_1\perp\mathbf{n}_2; \mathbf{n}_1\perp\mathbf{q};\,\mathbf{n}_2\perp\mathbf{q}$). $L_D$ is proportional to the ratio of the counter-acting exchange and Dzyaloshinskii constants and introduces a fundamental \textit{length} characterizing the periodicity of chiral modulations in chiral magnets, $L_D=A/D$.

The helical modulations have a fixed rotation sense determined by the sign of Dzyaloshinskii-Moriya constant $D$ and are continuously degenerate with respect to propagation directions of the helical modulations in space.
An applied magnetic field lifts this degeneracy and stabilizes two types of one-dimensional modulations: cones and helicoids with propagation directions parallel and perpendicular  to the applied  magnetic field respectively. 

For helicoids, analytical solutions for the polar angle $\theta(x)$ of the magnetization written in spherical coordinates, $\mathbf{M} = M_s\left( \sin \theta(x) \cos \psi, \sin \theta(x) \sin \psi, \cos \theta(x) \right)$, 
are derived by solving a \textit{pendulum} equation $A d^2\theta/d x^2 -H\cos \theta =0$.
Such solutions are expressed as a set of elliptical functions \cite{Dz64} and describe a gradual expansion of the helicoid period with increased magnetic field  (see the set of angular profiles $\theta(x)$, e.g.,  in Fig. 4.1. of Ref. \onlinecite{thesis}).
In a sufficiently high magnetic field $H_H$  the helicoid  infinitely expands and transforms into a system of isolated non-interacting 2$\pi$-domain walls (kinks) separating domains with the magnetization along the applied field \cite{Dz64,JMMM94}. The dimensionless value of this critical field is $h_H=H_H/H_D=\pi^2/8=0.30843$ with $H_D=D^2/AM$. The azimuthal angle $\psi$, on the contrary, is fixed by the
different forms of the Lifshitz invariants. In particular $\psi=\pi/2$ for Bloch helicoids in cubic helimagnets and $\psi=0$ for Neel helicoids or cycloids in polar skyrmion hosts with C$_{nv}$ symmetry such as the lacunar spinels GaV$_4$S$_8$ \cite{Kezsmarki15,Ruff15}, GaV4Se8 \cite{Butykai17}, GaMo4S8 \cite{Butykai19} or VOSe2O5 \cite{Kurumaji17}. In this latter  class of materials, the polar axial symmetry gives rise to Lifshitz invariants different from that of the cubic helimagnets (Eq. (\ref{DMdens1})) and the modulation vectors are restricted to the plane perpendicular to the high-symmetry axis.

The conical state combines the properties of the ferromagnetic and the helical states as a compromise between the Zeeman and DM energies. 
The conical spirals retain their single-harmonic character with $\psi = z/2L_D$ and $\quad \cos \theta = |\mathbf{H}|/2H_D$.
The magnetization component along the applied field has a fixed value $M_{\bot} = M \cos \theta = MH/2H_D$ and the magnetization vector $\mathbf{M}$ rotates within the surface of a cone. The critical value $h_d=2H_D$ marks the saturation field of the conical state.
For the functional $W_0(\mathbf{M})$ in Eq. (\ref{DMdens1}) the conical phase is the global minimum over the whole range of applied fields where modulated states exist ($ 0 < h < h_d$), whereas helicoids and skyrmion lattices can exist only as metastable states. 

\begin{figure*}
\includegraphics[width=1.7\columnwidth]{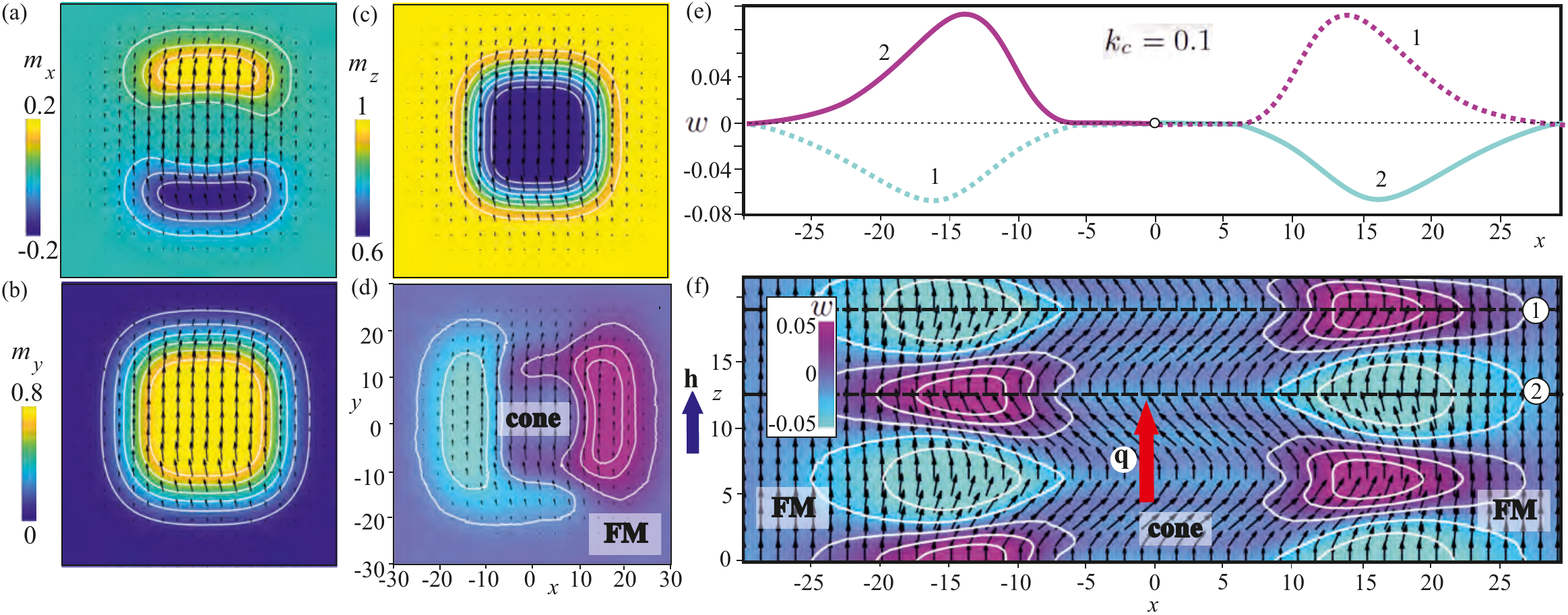}
\caption{ 
(color online) Internal structure of conical droplets formed during the first-order phase transition from the conical to the ferromagnetic state shown as color plots of the components of the magnetization (a)-(c) and  the  energy density $w$ (d).  The black arrows in (a)-(d) are projections of the magnetization on to the $xy$ plane, i.e. the same for all four panels. The $\mathbf{q}$-vector of the conical phase as well as the applied magnetic field $\mathbf{h}$ are directed along $z$. The energy density in the $xz$-cross-section of the conical droplet (f) exhibits alternating parts with the negative and positive contributions corresponding to "correct" and "wrong" senses of the magnetization rotation at the boundary to the ferromagnetic state. (e) shows the corresponding linear cross-sections along $x$ marked by the dashed lines 1 and 2 in (f). 
\label{Fig3} 
}
\end{figure*}

\section{Spiral states in helimagnets with cubic anisotropy}

In the following we discusse the influence of the cubic anisotropy  on the one-dimensional modulated states.
$W_a(\mathbf{m})$ in Eq. (\ref{DMdens1})  can be written as: 
\begin{equation}
W_a(\mathbf{m})= 
K_c (m_x^2m_y^2+m_x^2m_z^2+m_y^2m_z^2) 
\label{additional}
\end{equation}
where  $K_c$ is the coefficient of the cubic magnetic anisotropy
%
and the vector field of the magnetization is given by $\mathbf{m}(\mathbf{r}) = \mathbf{M}(\mathbf{r})/M_s$ under the constraint $|\mathbf{m}| = 1$. 

Generically, there are only small energy differences between various modulated states  including skyrmions. On the other hand, weaker energy contributions, such as $W_a(\mathbf{m})$, impose distortions on the solutions of the model (\ref{DMdens1}) which reflect the crystallographic symmetry and the values of magnetic interactions of the different chiral magnets. Thus, it is essential to recognize that these weaker interactions play a crucial role in determining the stability limits of the different modulated states on the corresponding phase diagrams.

In the case of single-ion cubic anisotropy,  the energy of the conical phase can be either increased or decreased depending on the mutual arrangement of the easy anisotropy axes and the propagation direction of the cones. 
%
A good starting point is 
a simple analysis of the homogeneous state in a system with the easy cubic anisotropy axis parallel to  the applied magnetic field as described in Ref. \onlinecite{JAP} and  shown in Fig. \ref{Fig2}.
In this case, the manifold of energy extrema provides a hint on the distortions of the conical state. The rotation of the magnetization becomes in tune with a complex landscape of the energy term associated, with the cubic anisotropy (Fig. \ref{Fig1}). 
The direction of the wave vector $\mathbf{q}$ itself is also specified by the competing Zeeman and cubic anisotropy energies.
These basic concepts are clearly manifested, in particular,  in the critical fields at the standard phase diagrams of bulk cubic helimagnets such as MnSi and Cu$_2$OSeO$_3$: the critical fields $H_{c1}$, where oblique spirals align along the field, and $H_{c2}$, where the conical phase undergoes a phase transition to the ferromagnetic state.

\subsection{ The conical-to-ferromagnetic phase transition at the critical field $H_{c2}$.}

\begin{figure*}
\includegraphics[width=1.7\columnwidth]{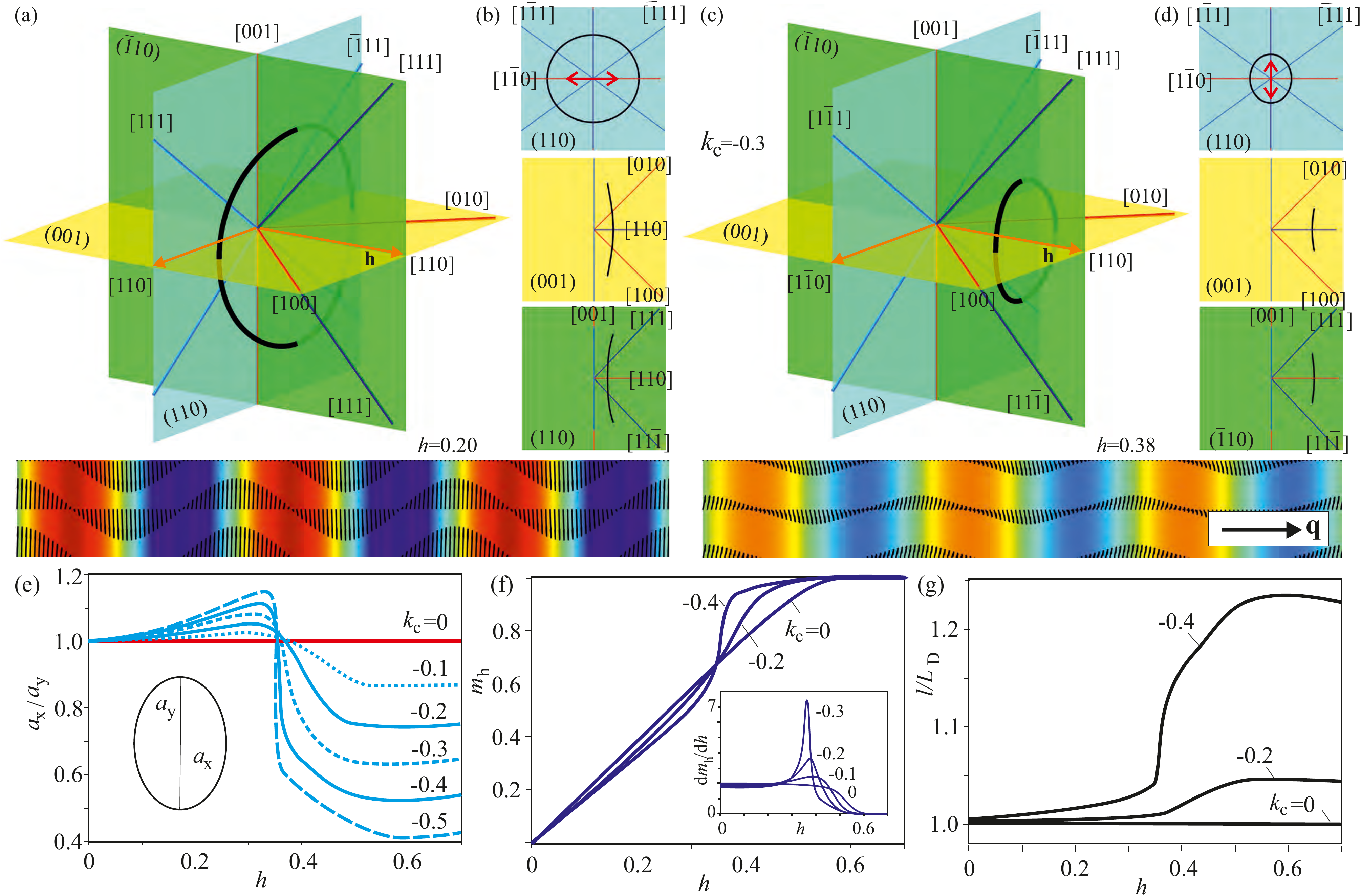}
\caption{ 
(color online)  $\mathbf{h}||[110]$, $k_c =-0.3$. Trajectories of the magnetization rotation in the conical phase (a), (c) in the presence of cubic anisotropy are depicted by the black curves for two values of the field, $h=0.2$ (a), $h=0.38$ (c). The anisotropy axes are represented by the red straight lines for the $\langle100\rangle$-type axes, and the blue lines for the $\langle111\rangle$-type axes. The corresponding projections of the magnetization traces on the planes (110), (001), ($\overline{1}$10) are  shown in (b) and (d), correspondingly, and reveal the change of ellipticity, which is also characterized by the ratio of ellipse half-axes in (e). Such an ellipticity change is accompanied by  magnetization jumps  and peaks in the magnetic susceptibility, as  shown in (f) and its inset.  For small anisotropy constants the periodicity changes only by a few percent (g). 
\label{Fig4} 
}
\end{figure*}

In this section we present solutions for spiral states, which we obtained by exact numerical minimization of the energy functional, given by Eq. (\ref{DMdens1}), including the cubic anisotropy $K_c$ (\ref{additional}). 
For this purpose, we consider the dimensionless anisotropy constant, $k_c = K_cA/D^2$. The magnetic field $\mathbf{H}$ is applied along the high-symmetry directions [110], [111], and [001].
The energy density is measured in non-dimensional units, $w(\mathbf{m})=W(\mathbf{M})/H_DM$.

\textit{a. First- or second-order phase transitions between the conical and the ferromagnetic state.}

In the present section, we consider the solutions for the conical state in the field applied along [001] or [111] for both signs of $k_c$. 
As it will become evident later, first-order phase transitions  occur when the magnetic field is applied along the easy anisotropy axes. In this case, on its way to saturation, the magnetization  is forced to embrace hard anisotropy axes, as shown in Fig. \ref{Fig1} (a) and (c). This is an energetically unfavourable configuration, that forces a discontinuous transition of the magnetisation towards the magnetic field. On the contrary, if the magnetic field points towards a hard anisotropy axis, the minima of the cubic anisotropy gradually align along the field and underlie the continuous second-order phase transition between the conical and the ferromagnetic state, as illustrated by Fig. \ref{Fig1} (b) and (c).
%
%

In the following, we carry out a comprehensive analysis of the magnetization rotation in the conical phase by considering  the surface energy plots shown in first row of Fig. \ref{Fig2}, and the three-dimensional polar plots of the cubic anisotropy energy depicted in the second row of Figs. \ref{Fig2}. The different topologies of these surface plots underly the particular types of the phase transitions.

For $k_c>0$ and $h=0$ (Fig. \ref{Fig2} (a)) the equilibrium states of the homogeneous state correspond to the easy axes of cubic anisotropy oriented along the $\langle001\rangle$ crystallographic directions. Maxima of the cubic anisotropy energy correspond to the hard $\langle111\rangle$ directions. 
Let us consider the helical mono-domain state where the magnetization for $h=0$ rotates in the plane (001). 
While rotating, the magnetization leaves one  energy minimum corresponding to $\langle001\rangle$ directions and, rotating  through the saddle point between hard axes $\langle111\rangle$, gets into another energy minimum along a $\langle001\rangle$ direction. The trace of the magnetization in the conical phase is shown by the yellow line in Fig. \ref{Fig2} .  
When applying a magnetic field along the [001] axis, the magnetization in the conical phase is forced to rotate in the vicinity of hard $\langle111\rangle$ axes, as shown in Fig. \ref{Fig2} (b) for $h=0.2$. This underlies the subsequent first-order phase into the ferromagnetic state,  which becomes evident, by e.g. the magnetization jumps in the magnetization curves shown in Fig. \ref{Fig1} (e). In Fig. \ref{Fig1} (d)  we plot the difference between the energy densities of the conical spiral and the ferromagnetic states. In this plot the point $B$ for $k_c=0.1$ stands for the first-order phase transition that takes place when the energy difference between the two phases vanishes.
Logically the same situation should also occur for $h||[111]$ and $k_c<0$, where three hard anisotropy axes $\langle001\rangle$ impose distortions to the conical phase and underlie the magnetization jump into the ferromagnetic state.

For $k_c<0$ and $h=0$ (Fig. \ref{Fig2} (c), (d)), the equilibrium states of the cubic anisotropy energy correspond  to the $\langle111\rangle$-type crystallographic directions. Thus, the maxima of the cubic anisotropy are the hard $\langle001\rangle$-type directions, and if a magnetic field is  applied along a  $\langle001\rangle$ axis, the angle between the easy directions and the field is $70.6^{\circ}$. 
In this case, the magnetization in the conical phase rotates so that it sweeps the easy directions $\langle111\rangle$ (Fig. \ref{Fig1} (b)). Even at zero field the magnetic structure is conical, thus, it has a non-zero uniform magnetization (see magnetization curves in Fig. \ref{Fig1} (g)). However, this  low-field conical phase is in reality a metastable solution overshadowed by the oblique spiral states with  $\mathbf{q}$-vectors along the easy $\langle111\rangle$-type directions.
In a finite magnetic field the global minima of the cubic anisotropy gradually approach the field direction and thus display a second-order phase transition to the ferromagnetic state (Fig. \ref{Fig2} (d)).
The same principle is valid also for $\mathbf{h}||[111]$ and $k_c>0$, where the angle between the field direction and the easy $\langle001\rangle$ axes is $54.7^\circ$, whereas for the hard $\langle111\rangle$ axes the angle is  $70.6^{\circ}$. In this case the topology of the corresponding  polar plots would  allow the gradual alignment of the magnetization along the hard  $\langle111\rangle$ direction with  increasing magnetic field.

\begin{figure*}
\includegraphics[width=1.99\columnwidth]{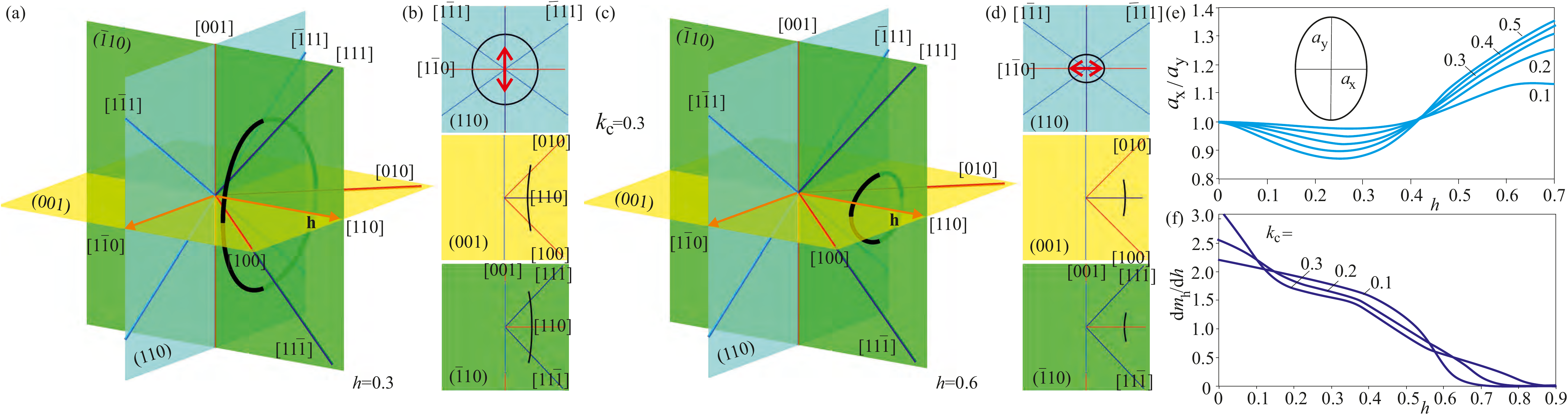}
\caption{ 
(color online) $\mathbf{h}||[110]$, $k_c = 0.3$. Trajectories of the magnetization rotation in the conical phase (a), (c) are shown for  $h=0.3$ (a), $h=0.6$ (c). 
The corresponding projections of the magnetization traces in (b) and (d) demonstrate a reverse ellipticity change as compared with Fig. \ref{Fig4}. This is also indicated by the ratio of ellipse half-axes in (e) showing first decrease and then a subsequent increase with the field. 
Interestingly, the characteristic fingerprints of the ellipticity change are not detected at the  magnetic susceptibility (f).
\label{Fig5} 
}
\end{figure*}

\textit{b. Conical domains in the ferromagnetic state.} 

As a consequence of the first-order phase transition that occurs for either $k_c>0$ and $h||[001]$ or $h||[111]$ and $k_c<0$, domains of the conical phase may persist in the field polarised state adopting the  configuration illustrated in Fig. \ref{Fig3}. 
Such domains exhibit both types of the rotational sense at their boundaries. 
At the left side of the depicted conical droplet (Fig. \ref{Fig3} (a)-(c)) the magnetization in the conical phase adjusts to the homogeneous state with the "correct" twist supported by DMI. This leads to energy gain as shown by the negative energy density difference in the $xy$ plane,  i.e., for a cross-section perpendicular to the magnetic field $\mathbf{h}$ and the conical $\mathbf{q}$-vector, shown in Fig. \ref{Fig3} (d)). At the right side of the conical droplet, the magnetization acquires the "wrong" in-plane twist, which results in an  energetic "penalty", as shown in Fig. \ref{Fig3} (d).

A similar behaviour is also found along the  $z$ direction, which is parallel to  $\mathbf{h}$ and $\mathbf{q}$. As illustrated in Fig. \ref{Fig3} (e), (f) the energy density of the conical droplet exhibits alternating negative and positive contributions corresponding to "correct" and "wrong" senses of the magnetization rotation at the boundary to the field polarised state. 
When the magnetic field is increased from the conical to the ferromagnetic phase,  the conical droplet may either continuously expand for $h < h_B$, as seen  in Fig. \ref{Fig1} (d) or gradually shrink for $h > h_B$,  a feature that underlies the kinetic character of such  first-order phase transition.
Remarkably, when  the magnetic field is decreased from the  field polarized to the conical state, the energy  density shown  in Fig. \ref{Fig3} (f) leads to an "in-phase" merge of neighboring conical droplets that initially may be formed with some phase shift with respect to each other. Then, the region with the positive energy density at the boundary of one conical droplet is annihilated by the corresponding part with the negative energy of another droplet, a feature that underlies in-phase "zipping" and "unzipping" of conical droplets. 

\begin{figure*}
\includegraphics[width=1.5\columnwidth]{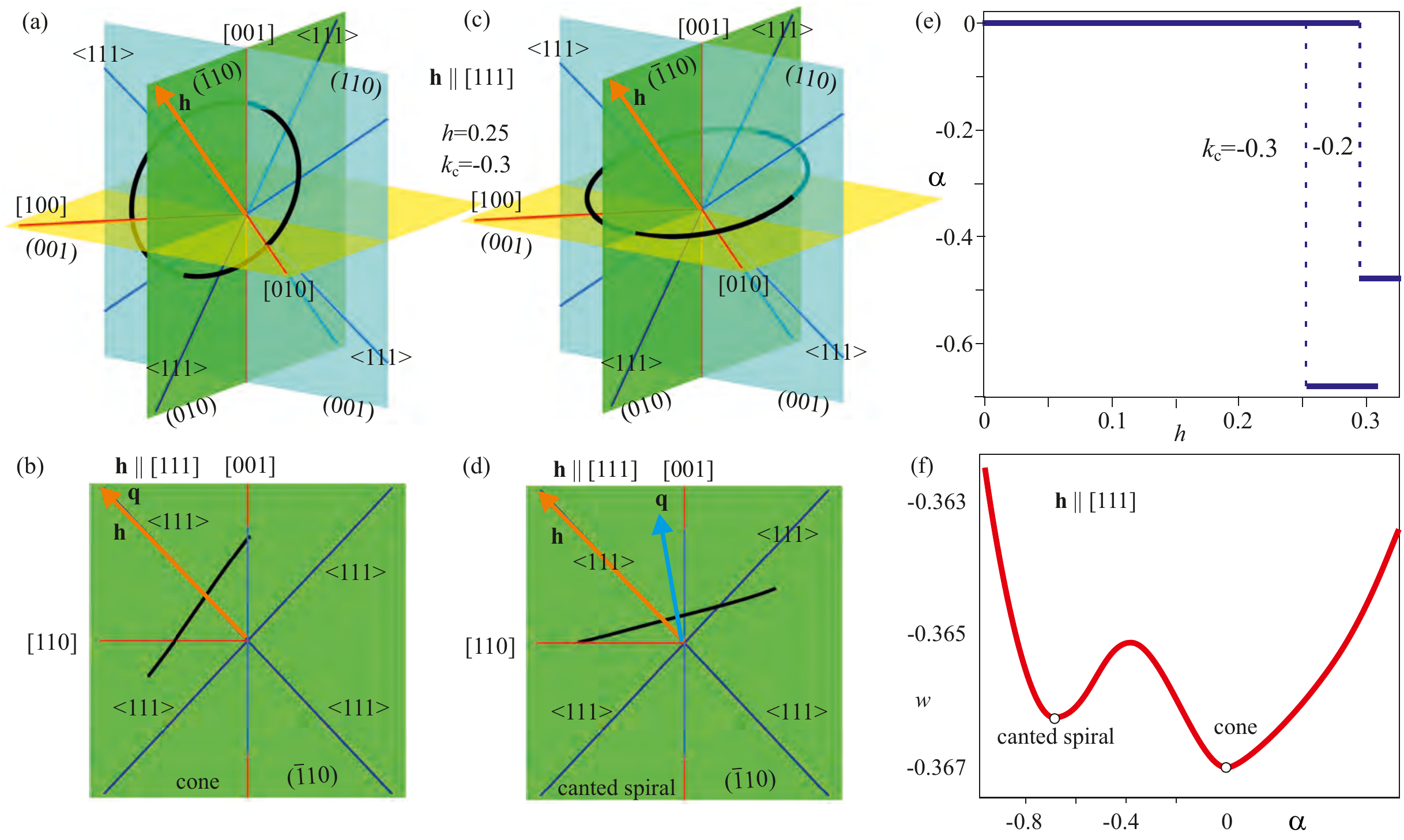}
\caption{ 
(color online)  Spiral reorientation near the saturation field $h_{c2}$,  for $\mathbf{h}||[111]$ and $k_c<0$. As seen from the sketches (a),(b), the magnetization in the conical state embraces two easy axes of the cubic anisitropy and is favored by the Zeeman energy. In the canted spiral state (c),(d), the magnetization  attempts to embrace all four easy $\langle111\rangle$ axes, at the expense of the Zeeman energy. This is a first-order phase transition as seen from the energy density (f), for $\mathbf{q}$ varied in the plane (010). The deflection angle $\alpha$ of the wave vector $\mathbf{q}$ with respect to the field, shown in (e),  depends on the value of $k_c$ and tends to $70.6^\circ$, thus to $\mathbf{q} \| [001]$ in the present configuration.  
\label{Fig6} 
}
\end{figure*}

\subsection{The helical-to-conical phase transition at the critical field $H_{c1}$}
\label{Hc1_section}

In this section we present qualitative results on the spiral reorientation processes that occur for $k_c<0$ 
 and for different directions of the applied magnetic field (Fig. \ref{Fig2} (e)).
Numerically exact solutions for the spiral states are obtained by direct minimization of the energy functional, given by Eq. (\ref{DMdens1}), including  $K_c$ in Eq. (\ref{additional}), and are discussed elsewhere \cite{chapter}. 
Since $k_c>0$ favours easy axes of $\langle111\rangle$ type, the ground state at zero field consists of domains of spiral states with the corresponding directions of the $\mathbf{q}$-vectors. 
Fig. \ref{Fig2} (d) shows a schematic layout of the $\mathbf{q}$-vectors in stable and metastable spiral states with respect to the field and the crystallographic directions.

For $\mathbf{H}\|[001]$ (first panel in Fig. \ref{Fig2} (e)), all four equivalent spiral states with  $\mathbf q\| \langle111\rangle$ are at an angle of $35.2^{\circ}$ with respect to the field. As also discussed in Ref. \onlinecite{chapter}, with increasing magnetic field, these wave vectors  first rotate gradually and then jump by a first-order phase transition towards the  field direction. 

For $\mathbf{H}\|[111]$ (second panel in Fig. \ref{Fig2} (e)), only the spiral that is  co-aligned with the magnetic field  is stable, whereas  the spirals along the other equivalent wave vectors $\mathbf q\| \langle111\rangle$  become metastable.
With  decreasing magnetic field only one spiral (or conical) domain with the $\mathbf{q}$-vector along  the field  persists and at zero field the other spiral domains  are not recovered.

For $\mathbf{H}\|[110]$ (third panel in Fig. \ref{Fig2} (e)), there are two sets of $\langle111\rangle$ directions composing smaller and larger angles with respect to the field. 
Thus the field-driven transition into the conical phase is a two-step process: the energetically unfavorable spiral states with the angle $90^{\circ}$ reorient first, whereas the stable spirals with the angle $27.35^{\circ}$  gradually rotate and eventually flip along the field. 
With decreasing magnetic field the transitions at both critical fields should occur in a reproducible manner, possibly with  some hysteresis.

\subsection{Phase transition between two elliptically distorted conical states for $\mathbf{H}||[110]$.}

For $\mathbf{H}||[110]$, independently on the sign of the cubic anisotropy constant $k_c$, the $\mathbf{q}$-vector of the conical state is surrounded by  two hard and two easy anisotropy axes $\langle111\rangle$ and $\langle100\rangle$.  This rosette of anisotropy axes leads to the characteristic elliptical distortion of the conical state, which moreover changes when the magnetization rotation occurs in the exterior or interior of the rosette. In particular for $k_c<0$ and low field values, the trajectory of the rotating magnetization is elongated in-plane along the hard $\langle100\rangle$ axes (Fig. \ref{Fig4} (a), (b)). With increasing magnetic field the magnetization approaches the rosette,  penetrates the rosette's interior and changes its elliptical distortion being now elongated along the easy $\langle111\rangle$ axes (Fig. \ref{Fig4} (c), (d)).

For $k_c>0$, the same process takes place, but with inverse switch of the elliptical distortion (Fig. \ref{Fig5}). The ratios of the ellipse major and minor axes for both signs of the anisotropy constant are plotted as a function of the magnetic field for different values of $k_c$ in Fig. \ref{Fig4} (e) and Fig. \ref{Fig5} (e). From these curves as well as from the magnetization curves (Fig. \ref{Fig4} (f)), it becomes clear that the phase transition between these two conical phases is more abrupt for the negative values of $k_c$. This might be explained by the  different angles the easy and hard anisotropy axes compose with respect to the field -- $35.3^{\circ}$ for $\langle111\rangle$ and $45^{\circ}$ for $\langle100\rangle$. Indeed, when $k_c<0$, the rotating magnetization tries to avoid two hard axes [100] and [010] and this results in the jump of the magnetization from the rosette's exterior to its interior. For $k_c>0$  [100] and [010] axes are easy anisotropy directions. Thus, although the sense of ellipticity changes, no jump of the magnetization is observed. 
An important point is that  the phase transition between the two different conical states leaves the periodicity of the spiral state almost unaffected. In fact, the period changes just by a few percent for low values of the cubic anisotropy (less than $4\%$ in Fig. \ref{Fig4} (g) for $k_c=-0.2$), and this change is hardly detectable by e.g. neutron scattering experiments.   
The field dependence of $dm_h/dh$ (Figs. \ref{Fig4} inset of (f) and Fig. \ref{Fig5} (f)), on the contrary, provides a  clear hint for this transition as it shows a $\lambda$-shaped spike only for $k_c<0$. 

\begin{figure*}
\includegraphics[width=1.5\columnwidth]{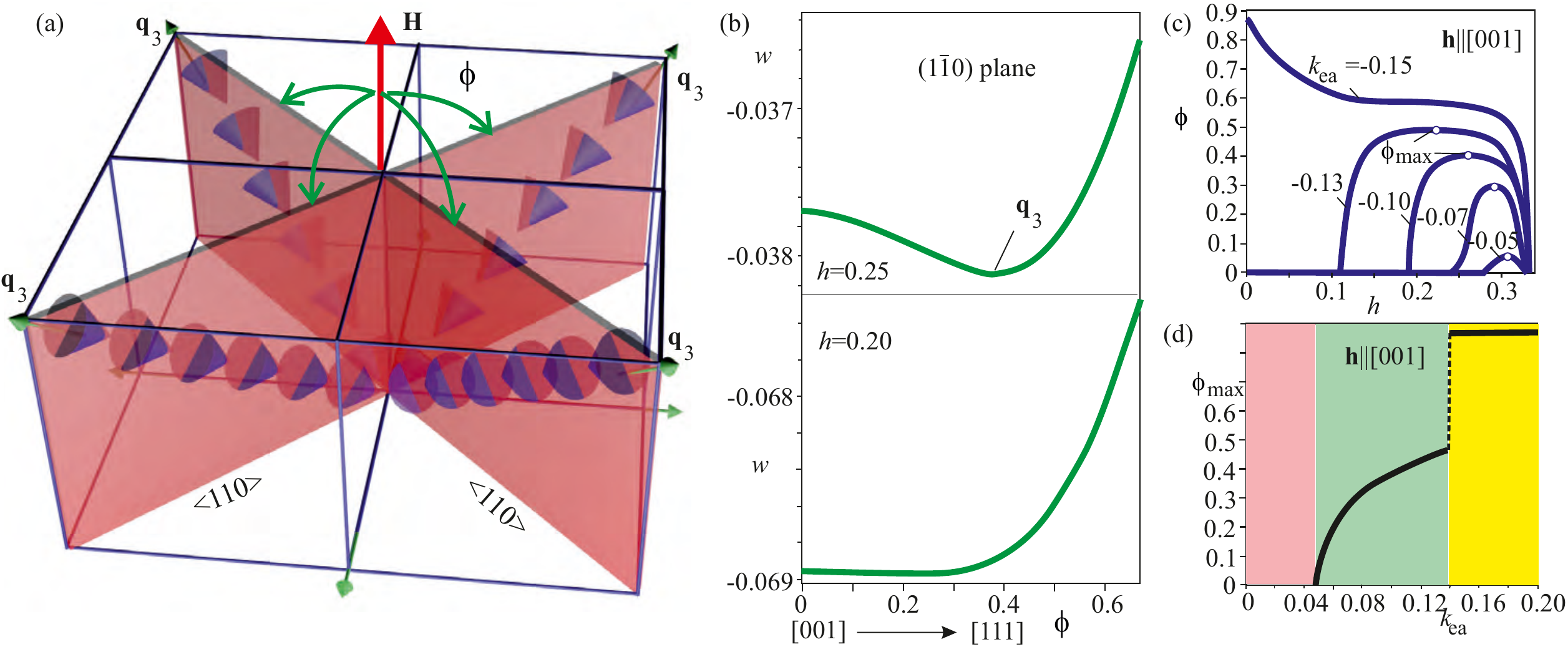}
\caption{ 
(color online) Schematics of the tilted-spiral state that appears above a critical value of the magnetic field, in which the wave vector $\mathbf{q}$ rotates away from the magnetic field $\mathbf{h}||[001]$ towards the $\langle111\rangle$ directions. (b) Energy density plotted as a function of the tilt angle $\phi$ in the (110) plane for $h = 0.2$ (lower panel) and $h=0.25$ (upper panel) exhibits the minimum shift. 
(c) Magnetic field dependence of the angle $\phi$  for different values of $k_{ea}$. Above the critical value, $k_{ea}\approx0.05$, $\phi$ increases from 0, which corresponds to the conical spiral phase, to a maximal value, $\phi_{max}$ (empty circle), and then decreases back to 0 as the magnetic field strength, $h$, increases. For $k_{ea}>0.14$, $\mathbf{q}||\langle111\rangle$  even at zero field and smoothly rotates toward [001] with increasing field.
(c) Dependence of  the maximal tilt angle $\phi_{max}$. For $k_{ea}<0.05$ (red area), $\phi_{max}=0$, which implies that $\mathbf{q}||[001]$. By increasing the strength of the anisotropic exchange, $\mathbf{q}$ smoothly rotates toward $\langle111\rangle$ (green
area), and for $k_{ea}>0.14$, $\mathbf{q}||\langle111\rangle$ (yellow area).
\label{Fig7} 
}
\end{figure*}

\subsection{ Spiral flip near the saturation field $H_{c2}$.}

In the following we will show that spiral reorientation processes akin to those at $H_{c1}$ may intervene also near the saturation field $H_{c2}$. 
Jumps of the spiral $\mathbf{q}$-vectors are less expected although they may be realized for relatively moderate values of the cubic anisotropy constant (Fig. \ref{Fig6}).
Indeed, the conical phase co-aligned with the field is considered to be the most favorable state that benefits from the Zeeman interaction. 
In the following, we show theoretically that a tilted spiral state may originate from the interplay between Zeeman and cubic anisotropy interactions, which is generic to chiral magnets, and thus may be important for explaining the experimental data of specific cubic helimagnets.
The spiral jumps are also sensitive to the direction of the applied magnetic field and the value of $k_c$, since relatively large values of $k_c$ are needed to implement particular reorientations. 

As an illustrative example, we consider the case with $\mathbf{H}||[111]$ and easy $\langle111\rangle$ axes of the cubic anisotropy (Fig. \ref{Fig6}). The $\mathbf{q}$-vector undergoes a first-order transition from the conical state (Fig. \ref{Fig6} (a),(b)) into a tilted spiral state with almost constant tilt angle for a range of magnetic field values  (Fig. \ref{Fig6} (c),(d)), with both the tilt angle and the field range dependent on $k_c$. The sketches in Fig. \ref{Fig6} (c)-(d) are instructive to deduce the nature of this tilted spiral state. Indeed, in an increasing magnetic field, the magnetization in the conical state sweeps only two $\langle111\rangle$ anisotropy axes (Fig. \ref{Fig6} (a), (b)). In tilted state, however, by sacrificing the Zeeman energy, the magnetization may try to embrace all four $\langle111\rangle$ axes (Fig. \ref{Fig6} (c), (d)). This leads to the maximum tilt angle $70.6^\circ$, reached for the higher $k_c$ values  (Fig. \ref{Fig6} (e)). 
This phase transition is accompanied by characteristic jumps in the magnetization curve. 
An interesting point is that such a reorientation phase transition requires a threshold value of $k_c$. For $k_c=-0.1$ not even a hint of the metastable tilted spiral becomes apparent in the angle-dependent energy density (like that in Fig. \ref{Fig6} (f)).

Conceptually the same reorientation process may occur also for $\mathbf{H}||[110]$. 

\section{Oblique spiral states due to the competing anisotropy interactions}

Tilted spiral states also may  originate from the interplay between competing anisotropic spin interactions, which is generic to chiral magnets. In fact, such an interplay gives a  rather good control over oblique states with smooth variation of the tilt angle.
As an instructive example, we consider spiral reorientation processes due to the competition between cubic anisotropy $k_c$ with the easy axes $\langle100\rangle$ and exchange anisotropy $k_{ea}<0$, i.e. with  easy $\langle111\rangle$ axes. 
This exchange anisotropy is defined as $w_{ea}=k_{ea}\left(\frac{\partial m_i}{\partial x_i} \right)^2$.

For the present setup of easy anisotropy axes, the tilted spiral state is found only for $\mathbf{h}||[001]$.
Above a critical field value, the conical spiral  (Fig. \ref{Fig1} (a), Fig. \ref{Fig2} (b))  begins to tilt towards one of the four body diagonals, the $\langle111\rangle$ directions, as shown in Fig. \ref{Fig7} (a).
Such a  tilted spiral state appears when  $|k_{ea}|$ exceeds a critical value, which is slightly lower than 0.05 for $k_c = 0.1$. 
Fig.~\ref{Fig7} (b) shows the dependence of the spiral energy on the tilt angle $\phi$ with {\bf q} varying in the (1$\bar 1$0) plane, for $h = 0.2$ and $h = 0.25$. 
As $h$ increases, $\phi$ grows, reaches its maximal value and then decreases back to zero, corresponding to a return into the conical spiral state (Fig. \ref{Fig7} (c)).  
The maximal tilt angle, $\phi_{\rm max}$, which  depends on the ratio of the competing fourth-order and exchange anisotropies, is plotted in Fig.~\ref{Fig7} (d).
As the  exchange anisotropy increases above the critical value, $|k_{ea}| > 0.14$,  the state with $\mathbf{q}\| \langle111\rangle$ is stabilised even  at zero magnetic field.

This effect can be understood as follows: the competing anisotropies give rise to  new minima of the conical spiral energy in the $\mathbf{q}$-space.
If only the fourth-order anisotropy $k_c$ is taken into account, the global energy minimum for {\bf H}$ \| \langle 001 \rangle$ is obtained for {\bf q}$ \| ${\bf H}. 
However, the additional anisotropy  term in Eq. (\ref{additional}) with an appropriate sign of $k_{ea}$ shifts these minima away from [001] towards the four directions $\langle111\rangle$ and thus underlies an instability of the conical state (Fig. \ref{Fig7} (a)).  

\section{Discussion and Conclusions}

To conclude, we have considered the magnetic field-driven reorientation transitions of spiral states that occur at the critical fields H$_{c1}$ and H$_{c2}$. We have shown that the theoretical model (1) predicts richer phase diagrams that are much more complex than assumed so far. Novel phases are stabilised, such as the an oblique spiral state that has recently been reported for the bulk helimagnet Cu$_2$OSeO$_3$ \cite{qian2018,chacon2018}. Moreover, phase transitions related to the ellipticity switch of spiral states may occur like the one that has been theoretically predicted for $H||[110]$ but has not been yet observed experimentally. Our theoretical findings can thus serve as a guideline for planning future experiments, that would systematically explore the role of impact of specific geometries of easy and hard anisotropy axes, on the ground states of chiral helimagnets. In the following we discuss the impact these exotic spiral states may have on the stabilisation and properties of chiral skyrmions.

\subsection{Spiral states and the internal structure of isolated skyrmions}

Remarkably, the internal properties of spiral states are directly imprinted on the functionalities of chiral skyrmions. 
Indeed, isolated skyrmions acquire a  magnetization distribution according to a surrounding "parental" or host  state.
In a state that is  homogeneously magnetized along the field, axially-symmetric skyrmion particles may form that underlie a repulsive inter-skyrmion potential \cite{LeonovNJP16}. 
On the other hand, when surrounded by a conical phase with the wave vector along the skyrmion axes, skyrmions develop a  three-dimensional  structure: the central core region nearly preserves the axial symmetry, whereas the domain-wall region, that connects the core with the embedding conical state acquires a crescent-like shape. 
This asymmetric profile of the cross-section forms a screw-like modulation along the skyrmion axis, matching the rotating magnetization of the conical phase at all cross-sections \cite{LeonovJPCM16}. 
To reduce the energy excess necessarily related to such a rotation, isolated skyrmions tend to form compact clusters with the unimpeded magnetization rotation in their interior.

Experimentally, clusters of  skyrmions have been observed in thin (70 nm) single-crystal samples of Cu$_2$OSeO$_3$  using transmission electron microscopy \cite{Loudon18}. 
In a bulk Cu$_2$OSeO$_3$ for $\mathbf{H}||[001]$, skyrmion clusters surrounded by the conical state exist up to the field $H_{c2}$ of the first-order phase transition between the cone and the FM state (Fig. \ref{Fig2} (a), (b)). 
With the onset of the homogeneous state, the inter-skyrmion attraction turns into repulsion and the constituent skyrmions disperse.
If the number of isolated skyrmions is sufficiently high, they may form a SkL with an equilibrium period that exists above the $H_{c2}$ line, as observed experimentally \cite{Bannenberg2019,chacon2018}.
However, if the number of skyrmions is low, they keep moving away from each other and the skyrmion signal in the experiment should fade away with time.
Thus, the following questions related to the skyrmion structure and skyrmion-skyrmion interaction arise: (i) What would be the impact of  the elliptically distorted conical states for $\mathbf{H}||[110]$ (Fig. \ref{Fig4}, \ref{Fig5}) or  the oblique spiral states (Fig. \ref{Fig7})? Does the skyrmion attraction become anisotropic in these cases? (ii) How does the structure of skyrmion clusters  distort during the spiral flips in Fig. \ref{Fig6}? Do skyrmions avert this sort of spiral jumps or are they instead annihilated via Bloch points? 


\subsection{Spiral states as  a background to accommodate differently oriented skyrmions}

Recently it was found that skyrmion tubes within the conical phase may orient either along or perpendicular to a wave vector \cite{Leonov2018a,Sohn19}. 
The first type of skyrmions  perfectly blends into the homogeneously saturated state after the cone closes in a strong magnetic field either by the first-order phase transition (Fig. \ref{Fig2} (b)) or by the second-order one (Fig. \ref{Fig2} (d)). 
The second type of skyrmions  perfectly blends into  the spiral state and represents two merons with equally distributed topological charge $Q = 1/2$.  
Then, a crossover between the two types of isolated skyrmions occurs for an intermediate value of an applied magnetic field. 
Thus, the found oblique spiral states (Fig. \ref{Fig6}, \ref{Fig7}) may underlie new skyrmion states oblique with respect to the field and their controllable switch by the field.

\subsection{Spiral states and the nucleation of skyrmions}

Skyrmion nucleation may occur via different mechanisms specified by the stable and metastable spiral states \cite{Bannenberg2019,droplets}. 
In particular, the low-temperature  stable SkL in the bulk-insulating cubic helimagnet Cu2OSeO3 for $\mathbf{H}||[001]$ may appear via rupture formation of spiral states with wave vectors perpendicular to the field \cite{Bannenberg2019},  along [100] and [010].
In this rather conventional mechanism also observed in the first papers on skyrmion visualization \cite{yuFeCoSi,YuFeGe}, the pairs of merons behave as if they were free particles. 
The properties of such skyrmions and merons within the spiral background were recently investigated in Refs. \onlinecite{Muller,Ezawa}.
The essential difference in this case is that the spiral states are  \textit{metastable} solutions, but give rise to a stable SkL. 

The experimental findings  in Cu${_2}$OSeO${_3}$\cite{Bannenberg2019} also hint that skyrmions may appear from the conical and the homogeneous states near the critical field $H_{cr2}$. This may occur via formation of magnetic torons of finite length accompanied by their elongation due to the cubic or exchange anisotropy, which  leads to the negative energy density of skyrmion cores and their mutual attraction. 
Such a homogeneous nucleation mechanism via torons -- spatially localized three-dimensional particles composed of a skyrmion filament of finite length cupped by two Bloch points terminating its prolongation -- was recently discussed in Refs. \onlinecite{leonov2018x,Mueller2020}.
In this sense, all sorts of the magnetization inhomogeneities provided, e.g., by the domain walls between spiral domains near the flip transition (Fig. \ref{Fig6}) or by the superstructure formed from the rotating oblique spiral (Fig. \ref{Fig7}), represent a favorable medium for the nucleation and stabbilisation of skyrmions. 

\textit{Acknowledgements. }
The authors are grateful to Christos Panagopoulos, Sandor Bordacs, Ulrich R\"o\ss ler, and Maxim Mostovoy for useful discussions. This work was funded by DFG Priority Program SPP2137, Skyrmionics, under Grant No. KE 2370/1-1. AOL thanks Ulrike Nitzsche for
technical assistance.

\end{document}